# Impact of epitaxial strain relaxation on ferromagnetism in a freestanding $La_{2/3}Sr_{1/3}MnO_3$ membrane

Ryuji Atsumi[1], Junichi Shiogai[1,2], Takumi Yamazaki[3], Takeshi Seki[3], Kohei Ueda[1,2], and Jobu Matsuno[1,2]

[1]*Department of Physics, Osaka University, Toyonaka, Osaka 560-0043, Japan*

[2]*Division of Spintronics Research Network, Institute for Open and Transdisciplinary Research Initiatives, Osaka University, Suita, Osaka 565-0871, Japan*

[3]*Institute for Materials Research, Tohoku University, Sendai 980-8577, Japan*

**Abstract**

Manganite perovskites host emerging physical properties of strongly-correlated electrons with charge, spin, and lattice degrees of freedom. Using epitaxial lift-off technique, we report enhancement of saturation magnetization and ferromagnetic transition temperature of the freestanding $La_{2/3}Sr_{1/3}MnO_3$ membrane compared with the as-grown film on $SrTiO_3$ substrate involving lateral tensile strain. Structural analysis reveals shrinkage of unit-cell volume by tensile strain relaxation in the freestanding membrane, which causes enhancement of the ferromagnetic interaction. The impact of the microscopic lattice deformation on the ferromagnetism of $La_{2/3}Sr_{1/3}MnO_3$ indicates a high potential of this material for flexible electronics application with intriguing functionalities in strongly-correlated electron systems.

The demand for preparation of thin films and heterostructures composed of various functional materials on flexible substrates is rapidly increasing owing to recent development of flexible electronics.[1,2] However, synthesis of highly crystalline thin film is not compatible to the flexible substrates made of organic polymers, because it usually requires high-temperature process and lattice matching. Therefore, tailoring functional materials in thin-film form on arbitrary substrates has been limited to mechanical exfoliation from bulk single crystal of van der Waals materials[3,4] or low-temperature deposition.[5,6] Recently, the epitaxial lift-off technique,[7,8] where the sacrificial underlayer is selectively removed by chemical etching, has been utilized as a powerful tool to transfer the highly crystalline films of various materials on various substrates. After the invention of a cubic $(Ca,Sr,Ba)_3Al_2O_6$ water-soluble sacrificial layer,[9,10,11] the synthesis and characterization of freestanding membranes of various functional oxides and chalcogenides have been particularly focused on, including $SrTiO_3$,[9,12] transparent conducting oxide,[10] high-temperature superconducting cuprate,[13] manganite perovskites,[9,14,15,16] and iron-based superconductor[17] on some substrate materials such as polymer and a thermally oxidized silicon.

Transition metal oxides exhibit a wide variety of exotic electronic and magnetic phases owing to competition between orbital, spin and charge degrees of freedom of electrons.[18,19] Among such material systems, manganite perovskites exhibit colossal magnetoresistance, half metallicity, and ferromagnetism with relatively high ferromagnetic-paramagnetic transition temperature ($T_C$), making this class of materials not only a research platform of strongly correlated physics, but also an attractive candidate constituent of magnetic and/or spintronic devices[20]. The key feature of the manganite perovskites is the doubly degenerated $e_g$ state on the Mn site in the $MnO_6$ octahedron; its degeneracy is lifted by biaxial strain and electron hopping between $e_g$ orbitals is modulated by Mn-Mn separation or tilting angle of Mn-O-Mn bonding between adjacent $MnO_6$ octahedrons. Therefore, electrical and magnetic properties should be sensitive to the lattice deformation. Actually, it has been demonstrated that tensile and compressive epitaxial strain from the lattice-mismatched substrates largely influence magnetic anisotropy,[21] electrical conductivity[22], and $T_C$ [Ref. 22,23] of epitaxial manganite thin films. It has been recently reported that a freestanding $(La,Ca)MnO_3$ membrane with $T_C$ of about 250 K, formed on a stretchable substrate by the epitaxial lift-off technique, exhibits a drastic change in electronic phases by application of extreme strain up to 8% without fractures,[14] which is inaccessible in ordinary epitaxial approach. More recently, it has been shown that a freestanding $(La,Sr)MnO_3$ membrane exhibits the room-temperature ferromagnetism[9,15,16], which has an advantage for practical application, and the relaxation of epitaxial strain during the epitaxial lift-off process is suggested to cause the modulation of ferromagnetism.[16] For implementation in flexible electronics, a microscopic understanding of link between lattice deformation and the room-temperature ferromagnetism in the freestanding $(La,Sr)MnO_3$ membrane is highly desirable.

In this study, we synthesis a 22-nm-thick room-temperature ferromagnetic $La_{2/3}Sr_{1/3}MnO_3$ membrane on a flexible substrate by epitaxial lift-off technique. From structural characterization and magnetization measurement, we observed an enhancement of saturation magnetization and a slight increase of $T_C$ in fully relaxed membrane by comparing with as-grown strained thin film on a $SrTiO_3$ substrate. A clear link of the strain and magnetic properties yields the potential application of the freestanding manganite membrane for flexible electronics based on strain-induced emergent electrical and magnetic functionalities.[24]

The La$_{2/3}$Sr$_{1/3}$MnO$_3$ membrane was fabricated by epitaxial lift-off technique as illustrated in Fig. 1(a). First, a heterostructure consisting of 60-nm-thick amorphous AlO$_x$/22-nm-thick La$_{2/3}$Sr$_{1/3}$MnO$_3$/22-nm-thick Sr$_3$Al$_2$O$_6$ was prepared on the single-crystalline SrTiO$_3$(001) substrate by pulsed laser deposition (PLD) using the stoichiometric targets. The Sr$_3$Al$_2$O$_6$ sacrificial layer was grown at substrate temperature $T_{sub}$ of 800°C under the oxygen partial pressure $P_{O2}$ of $1 \times 10^{-5}$ Torr and subsequently La$_{2/3}$Sr$_{1/3}$MnO$_3$ thin films was grown at $T_{sub}$ = 750°C under $P_{O2}$ = $2.5 \times 10^{-5}$ Torr. After the high-temperature growth of Sr$_3$Al$_2$O$_6$ and La$_{2/3}$Sr$_{1/3}$MnO$_3$, $T_{sub}$ was cooling down to room temperature and the AlO$_x$ layer was deposited as capping layer in vacuum. The AlO$_x$ capping layer is critical to suppress formation of cracks during lift-off process.[25] After the thin-film growth process, the as-grown AlO$_x$/La$_{2/3}$Sr$_{1/3}$MnO$_3$/Sr$_3$Al$_2$O$_6$ heterostructure was attached to a 94-μm-thick flexible and adhesive polyimide (PI) sheet (Kapton, DuPont Co.) and immersed in distilled water at room temperature for 24 hours. The dissolution of the Sr$_3$Al$_2$O$_6$ sacrificial layer in water results in exfoliation of the AlO$_x$ capped La$_{2/3}$Sr$_{1/3}$MnO$_3$ membrane from the substrate as shown in Fig. 1(b). Figure 1(c) shows optical micrograph of a AlO$_x$ capped La$_{2/3}$Sr$_{1/3}$MnO$_3$ after epitaxial lift-off process. The continuous membrane with the area of about $500 \times 500$ μm$^2$ was obtained. One can find that the size of membrane is relatively large by comparing with the uncapped La$_{2/3}$Sr$_{1/3}$MnO$_3$ membrane. The released PI tape/AlO$_x$/La$_{2/3}$Sr$_{1/3}$MnO$_3$ heterostructure membrane was then fixed onto a glass substrate for x-ray diffraction (XRD) measurement. Cross-sectional high-angle annular dark field (HAADF) scanning transmission electron microscope (STEM) images were taken for microscopic structural analysis and thickness evaluation. Measurements of magnetic hysteresis loop [$M(H)$ curve] at room temperature and temperature dependence of magnetization [$M(T)$ curve] were performed by room-temperature vibrating sample magnetometer (VSM, Tamakawa Co., Ltd) and temperature variable SQUID magnetometer (MPMS, Quantum Design Inc.), respectively.

Figure 2(a) shows the 2theta-omega XRD pattern of the as-grown $AlO_x$/$La_{2/3}Sr_{1/3}MnO_3$/$Sr_3Al_2O_6$ heterostructure around $La_{2/3}Sr_{1/3}MnO_3$(002) diffraction peak, clearly indicating growth of epitaxially oriented $La_{2/3}Sr_{1/3}MnO_3$ and $Sr_3Al_2O_6$. Hereafter, the pseudocubic notation is employed for representing crystallographic plane of $La_{2/3}Sr_{1/3}MnO_3$, and out-of-plane and in-plane lattice constants are denoted as $c$- and $a$-axes, respectively, for simplicity. The $c$-axis length of $Sr_3Al_2O_6$ is 15.856 Å, being close to the bulk value (15.844 Å). Clear thickness fringes around $La_{2/3}Sr_{1/3}MnO_3$(002) indicate a flat surface morphology. The $c$-axis length of $La_{2/3}Sr_{1/3}MnO_3$ thin film was determined to be $c$ = 3.833 Å from (002) diffraction peak, which is consistent with the previously reported value of the $La_{2/3}Sr_{1/3}MnO_3$ thin film with the comparable thickness grown on a $SrTiO_3$ substrate.[23] Figure 2(b) shows the 2theta-omega XRD pattern of the released $AlO_x$/$La_{2/3}Sr_{1/3}MnO_3$ membrane. The $La_{2/3}Sr_{1/3}MnO_3$ (002) peak is clearly observed, while any diffraction peaks from $SrTiO_3$ and $Sr_3Al_2O_6$ are not detected, revealing that the crystal orientation of $La_{2/3}Sr_{1/3}MnO_3$ is preserved and the $SrTiO_3$ substrate and the $Sr_3Al_2O_6$ layer were completely removed during the etching in water. As can be seen in Figs. 2(a) and 2(b), the $La_{2/3}Sr_{1/3}MnO_3$ (002) diffraction peak of the membrane slightly shifts from the film value to a smaller angle, corresponding to the elongated $c$-axis length to 3.844 Å [a vertical dashed line in Figs. 2(a) and 2(b)]. Assuming that the released $La_{2/3}Sr_{1/3}MnO_3$ membrane is fully relaxed, the as-grown film is subjected to the compressive strain of $\varepsilon_{zz}$ = −0.29% along $c$ axis probably due to the in-plane biaxial tensile strain introduced from the $Sr_3Al_2O_6$ underlayer ($a$/4 = 3.964 Å).

For microscopic structural analysis, cross-sectional HAADF-STEM images were taken for the as-grown AlO$_x$/La$_{2/3}$Sr$_{1/3}$MnO$_3$/Sr$_3$Al$_2$O$_6$ heterostructure on the SrTiO$_3$ substrate and AlO$_x$/La$_{2/3}$Sr$_{1/3}$MnO$_3$ membrane on the PI substrate as shown in Figs. 2(c) and 2(d), respectively. For both images, the incident electron beam was set to be parallel to La$_{2/3}$Sr$_{1/3}$MnO$_3$[010]. The sharp interface between each layer in the as-grown heterostructure, consistent with the observation of fringe patten around La$_{2/3}$Sr$_{1/3}$MnO$_3$ XRD peak shown in Fig. 2(a). For the released membrane, the thicknesses of the AlO$_x$ and La$_{2/3}$Sr$_{1/3}$MnO$_3$ layer are unchanged while the Sr$_3$Al$_2$O$_6$ layer is completely removed as shown in Fig. 2(d), elaborating that water acts as selective etchant for the Sr$_3$Al$_2$O$_6$ layer. Figures 2(e) and 2(f) show the magnified HAADF-STEM image around the bottom interface of the La$_{2/3}$Sr$_{1/3}$MnO$_3$ layer and variation of its $a$-axis length along the growth direction for the as-grown heterostructure and membrane, respectively. The obtained $a$-axis length of the La$_{2/3}$Sr$_{1/3}$MnO$_3$ layer in the as-grown heterostructure gradually decreases from 3.94 Å at the interface with of the Sr$_3$Al$_2$O$_6$ underlayer to 3.86 Å toward the top surface. The variation of $a$-axis length indicates that the La$_{2/3}$Sr$_{1/3}$MnO$_3$ layer is subjected to the tensile strain from Sr$_3$Al$_2$O$_6$ with $a/4 = 3.964$ Å within the region of ~10 nm from the interface. For the membrane on the PI substrate, in contrast, a variation of the $a$-axis length is not obvious, and its value is around 3.86 Å, which is consistent with the value of the La$_{2/3}$Sr$_{1/3}$MnO$_3$ layer in the relaxed region in the as-grown heterostructure. In addition, the $a$-axis length of the relaxed membrane is comparable to its $c$-axis length (3.844 Å) determined by XRD. Based on the analysis using XRD and STEM, we conclude that the epitaxial strain of the as-grown La$_{2/3}$Sr$_{1/3}$MnO$_3$ film around the interface with the Sr$_3$Al$_2$O$_6$ underlayer corresponds to $\varepsilon_{xx} = 2.1\%$, which is fully relaxed in the membrane form.

The magnetic properties of the as-grown La$_{2/3}$Sr$_{1/3}$MnO$_3$ thin film on the $Sr_3Al_2O_6$ layer and the released $La_{2/3}Sr_{1/3}MnO_3$ membrane were first characterized by $M(H)$ curve at room temperature. Figures 3(a) and 3(b) show $M(H)$ curves for the as-grown thin film and membrane at room temperature when the magnetic field was swept along [100] crystallographic direction. The magnetic field was applied up to $\mu_0 H = 50$ and 100 mT, which is large enough to saturate the magnetization of the as-grown film and the released membrane, respectively. As can be seen in Figs. 3(a) and 3(b), the obtained value of saturation magnetization is 0.16 $\mu_B$/Mn for the as-grown film [in Fig. 3(a)] while that is 0.43 $\mu_B$/Mn for the released membrane [in Fig. 3(b)]. The observed difference in the saturation magnetization for as-grown film and membrane suggests the two possible mechanisms: modulation of $T_C$ and saturation magnetization at ground state, which is discussed based on $M(T)$ curve.

Figure 3(c) shows $M(T)$ curves for the as-grown film and the released membrane. During the measurement in descending temperature, the application of $\mu_0 H = 0.1$ T was needed to saturate magnetization of the $La_{2/3}Sr_{1/3}MnO_3$ layer. For evaluating the values of $T_C$ from $M(T)$ curve with a broad ferromagnetic transition behavior, we employed the inflection of $M(T)$ obtained by the temperature derivative of $M(T)$ [$dM(T)/dT$]. The inflection of $M(T)$ occurs at 305 K for membrane, while it occurs at 295 K for the as-grown film as seen in the bottom panel in Fig. 3(c), corresponding to 3% increase of $T_C$ by releasing the epitaxial strain. In addition, the saturation magnetization in overall temperature is clearly larger for the released membrane than that of the as-grown film. The saturation magnetization at $T = 5$ K reaches about 2.25 $\mu_B$/Mn in membrane form, which is 30% increase from the value of 1.74 $\mu_B$/Mn in the as-grown thin film. Therefore, the slight increase of $T_C$ and enhanced saturation magnetization at ground state play a role for the increase of saturation magnetization at room temperature. By considering mixed valence of Mn in the stoichiometric La$_{2/3}$Sr$_{1/3}$MnO$_3$, the ideal saturation magnetic moment should be 3.33 $\mu_B$/Mn. However, fully relaxed membrane reaches only 2.25 $\mu_B$/Mn at the lowest temperature, which can be ascribed to the off-stoichiometry of cation ratio or/and the presence of oxygen deficiency.[26]

Finally, we discuss microscopic origins for the modulation of $T_C$ and saturation magnetization. According to the structural analysis in Fig. 2, the as-grown thin film is subjected to in-plane tensile strain from the cubic crystal structure with $\varepsilon_{xx}$ = 2.1% and $\varepsilon_{zz}$ = −0.29% as determined by STEM and XRD, respectively. The estimated unit-cell volume is indeed enlarged by $2\varepsilon_{xx} + \varepsilon_{zz}$ ~ 4% for the as-grown thin film, which weakens the ferromagnetic interaction. In bulk single crystal of La$_{2/3}$Sr$_{1/3}$MnO$_3$, the $T_C$ modulation coefficient by hydrostatic pressure is reported to be 1-2%/GPa.[27] Using reported value of the room-temperature bulk modulus of La$_{2/3}$Sr$_{1/3}$MnO$_3$ of 167 GPa,[28] shrinkage of unit-cell volume by 4% should correspond to 6.7-13% increase of $T_C$. Although the observed change in $T_C$ of 3% is smaller than that expected, probably because the as-grown La$_{2/3}$Sr$_{1/3}$MnO$_3$ film is partly relaxed in ~10 nm apart from the bottom interface, the increased $T_C$ and saturation magnetization in the freestanding membrane can be semi-quantitatively explained by modulation of unit-cell volume.

In summary, we fabricated freestanding La$_{2/3}$Sr$_{1/3}$MnO$_3$ membrane by epitaxial lift-off technique using Sr$_3$Al$_2$O$_6$ as a water-soluble sacrificial layer. Both the as-grown film and freestanding membrane of La$_{2/3}$Sr$_{1/3}$MnO$_3$ exhibited a clear ferromagnetic behavior at room temperature. The saturation magnetization and ferromagnetic transition temperature for the freestanding membrane were enhanced from the value for the as-grown film. Based on the microscale structural analysis, we found that relaxation of in-plane tensile strain and shrinkage of the unit-cell volume in the membrane form from the as-grown La$_{2/3}$Sr$_{1/3}$MnO$_3$ film most likely to be the origin of modulation of ferromagnetism. The impact of microscopic lattice deformation on the room-temperature magnetic properties of La$_{2/3}$Sr$_{1/3}$MnO$_3$ indicates a high potential of this materials for flexible electronics application based on strain-engineered physical properties and functionalities with strongly-correlated electrons.

**Acknowledgments**

The authors thank Shun Ito and Kana Takenaka at Analytical Research Core for Advanced Materials, Institute for Materials Research, Tohoku University, for STEM observation, and Kensuke Takaki, Hirotake Suzaki, and Kazutaka Kudo at Osaka University for chemical analysis facility. This work was partly performed under the GIMRT Program of the Institute for Materials Research, Tohoku University (Proposal No. 202212-CRKEQ-0007). This work was supported by JST, PRESTO Grant Number JPMJPR21A8, Japan, JSPS KAKENHI Grant Number JP22K18894, and Iketani Science and Technology Foundation. T.Y. is supported by JSPS through a Research Fellowship for Young Scientists (Grant Number JP22KJ0210).

**Figure Captions**

**Fig. 1.** (a) Schematic illustration for releasing $AlO_x$ capped $La_{2/3}Sr_{1/3}MnO_3$ (LSMO) membrane from $SrTiO_3$ (STO) substrate by etching $Sr_3Al_2O_6$ (SAO) sacrificial layer. (b) Photograph of the LSMO freestanding membrane on a polyimide substrate. (c) Optical micrograph of the LSMO freestanding membrane after etching the sacrificial layer.

**Fig. 2.** (a)(b) 2theta-omega X-ray diffraction (XRD) patterns for (a) as-grown $AlO_x/La_{2/3}Sr_{1/3}MnO_3/Sr_3Al_2O_6$ heterostructure on $SrTiO_3$ substrate and (b) released $AlO_x/La_{2/3}Sr_{1/3}MnO_3$ membrane. (c)(d) Wide view of cross-sectional HAADF-STEM images for (c) as-grown and (d) released membrane. (e)(f) Magnified views [orange rectangle in (c)(d)] of cross-sectional HAADF-STEM images for (e) as-grown and (f) released membrane and corresponding line profiles of *a*-axis length of $La_{2/3}Sr_{1/3}MnO_3$ along the growth direction.

**Fig. 3** (a)(b) Magnetic hysteresis loops for (a) as-grown $La_{2/3}Sr_{1/3}MnO_3$ film on the $Sr_3Al_2O_6$ layer and (b) released $AlO_x/La_{2/3}Sr_{1/3}MnO_3$ membrane measured at room temperature. (c) (Top) Temperature dependence of magnetization $M(T)$ for as-grown film (green) and released membrane (blue). Measurement was performed at in-plane field of $\mu_0 H = 0.1$ T. (Bottom) Temperature dependence of $dM(T)/dT$ obtained from temperature-differentiated $M(T)$.

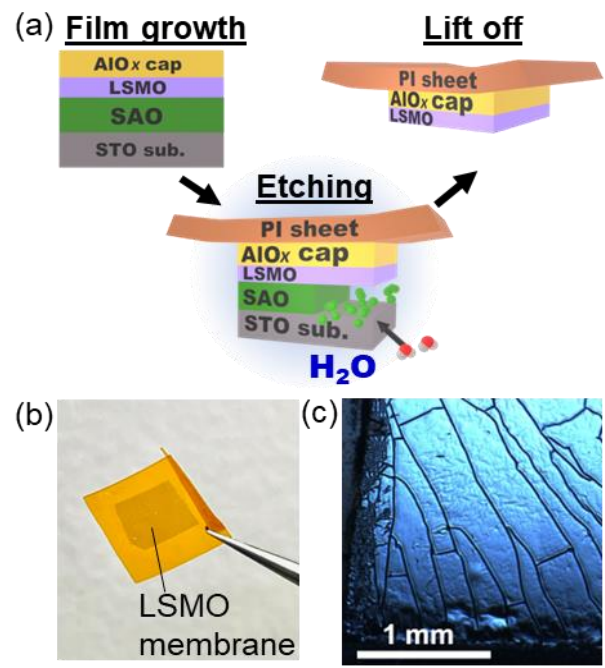

Fig.1. R. Atsumi et al.

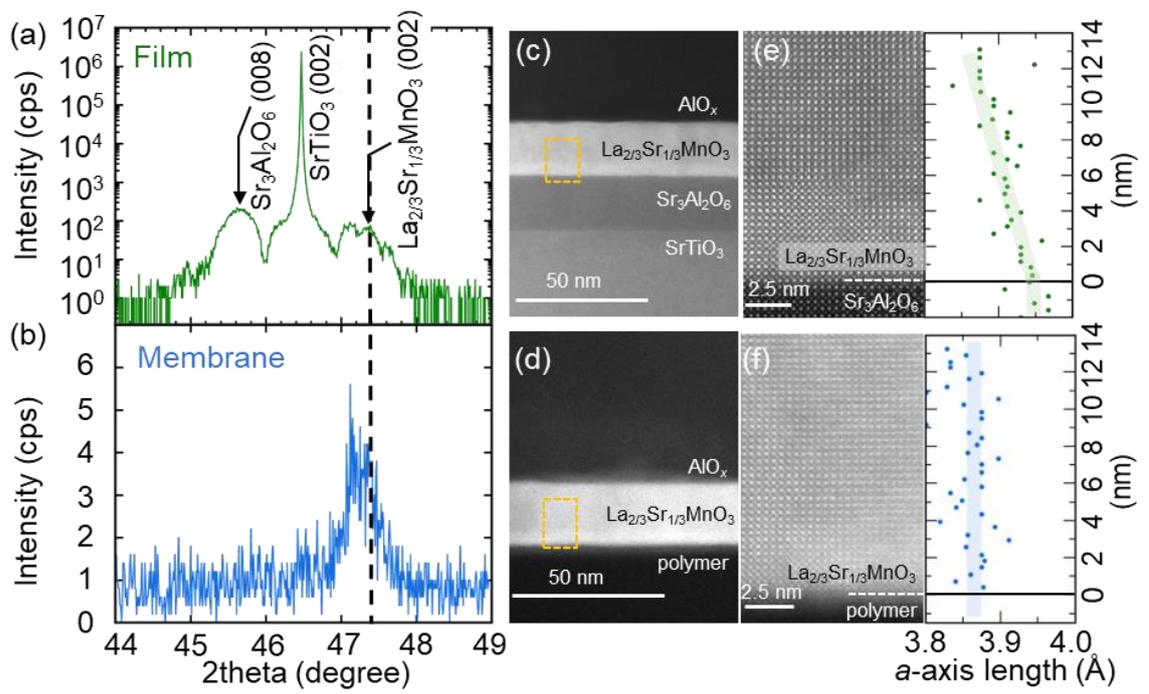

Fig. 2. R. Atsumi et al.

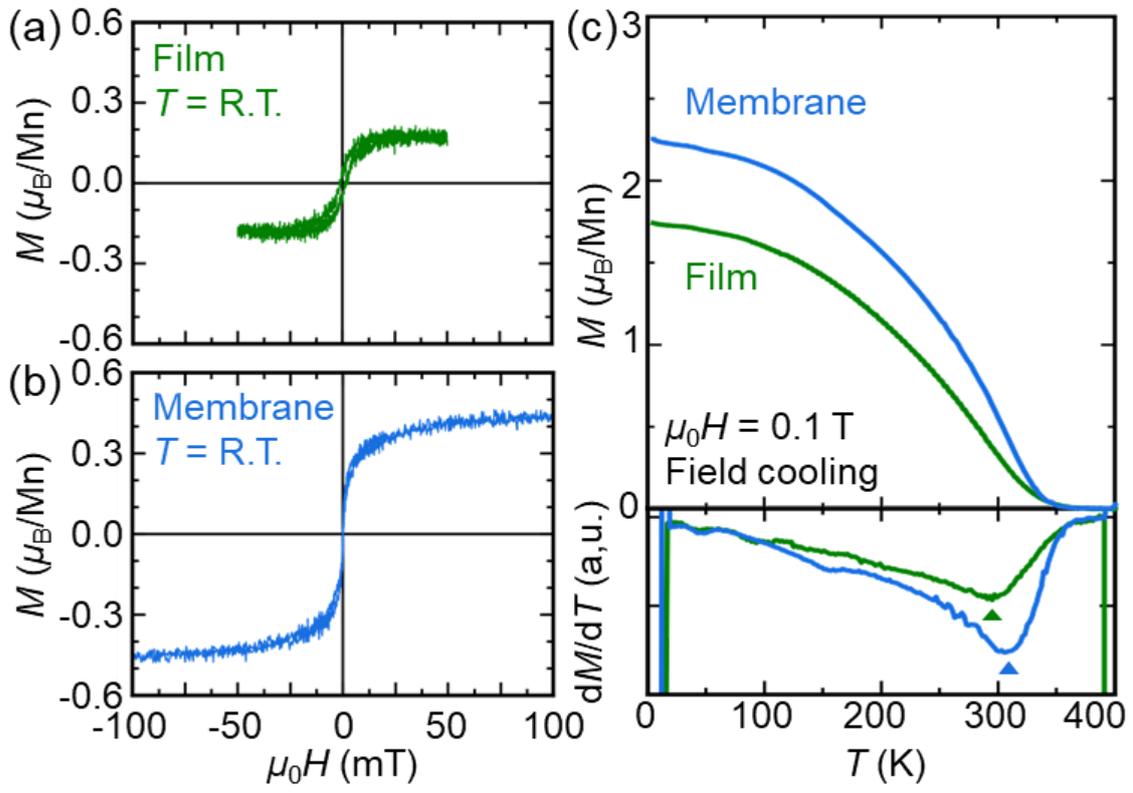

Fig. 3. R. Atsumi et al.